# A RESISTIVE FORCE MODEL FOR LEGGED LOCOMOTION ON GRANULAR MEDIA[*]

CHEN LI, TINGNAN ZHANG, AND DANIEL I. GOLDMAN

*School of Physics, Georgia Institute of Technology, 837 State St NW*
*Atlanta, GA 30332, USA*

Compared to agile legged animals, wheeled and tracked vehicles often suffer large performance loss on granular surfaces like sand and gravel. Understanding the mechanics of legged locomotion on granular media can aid the development of legged robots with improved mobility on granular surfaces; however, no general force model yet exists for granular media to predict ground reaction forces during complex limb intrusions. Inspired by a recent study of sand-swimming, we develop a resistive force model in the vertical plane for legged locomotion on granular media. We divide an intruder of complex morphology and kinematics, e.g., a bio-inspired robot L-leg rotated through uniform granular media (loosely packed ~ 1 mm diameter poppy seeds), into small segments, and measure stresses as a function of depth, orientation, and direction of motion using a model leg segment. Summation of segmental forces over the intruder predicts the net forces on both an L-leg and a reversed L-leg rotated through granular media with better accuracy than using simple one-dimensional penetration and drag force models. A multi-body dynamic simulation using the resistive force model predicts the speeds of a small legged robot (15 cm, 150 g) moving on granular media using both L-legs and reversed L-legs.

## 1. Introduction

A variety of desert animals including insects, lizards, and mammals encounter granular surfaces like sand and gravel in their natural environments and appear to move across them with ease [1]. In contrast, man-made vehicles, most of which move on wheels and tracks, can become stuck on granular media [2] because these substrates yield and flow under sufficient load. Understanding how legged animals generate sufficient ground reaction forces without sinking deeply on granular media will help improve robotic mobility on these challenging substrates [3]. However, unlike fluids for which forces can be computed by the Navier-Stokes equations [4], no comprehensive force model yet exists for granular media. While classical terramechanics has guided the design and performance tests of large, off-road, wheeled and tracked vehicles [5, 6], the empirical and computational methods were developed specifically for wheel/track-terrain interfaces and thus it is not clear if they apply to small legged locomotors [5, 6].

---

[*] This work was supported by The Burroughs Wellcome Fund, Army Research Lab MAST CTA, and the Army Research Office.

Despite recent progress in modeling limb-ground interactions during legged locomotion on granular media [1, 7, 8], there is need to develop a force model to accurately and quickly predict forces on intruders of arbitrary shapes and trajectories moving through granular media. These recent studies [1, 7, 8] modeled the ground reaction forces in granular media with one-dimensional force models, such as the force on a vertically penetrating disc oriented horizontally [9], or the force on a horizontally dragged plate oriented vertically [10]. While these studies reveal some basic mechanisms of locomotion on granular media, these models do not fully capture forces on intruding limbs that are complex in both morphology and kinematics [1, 7, 8, 11, 12]. The discrete element method (DEM) which calculates forces on all interacting particles and the intruder [13] is accurate in predicting forces, but it is computationally expensive and limited in the number of particles that can be simulated, thus restricting the size of the granular bed or the size of particles [13].

## 2. Development of resistive force model

Our resistive force model for legged locomotion was inspired by a resistive force model developed in the study of a small lizard swimming horizontally within granular media [14]. The resistive force model divides an intruder of complex morphology and kinematics into small segments, each generating forces that are assumed independent. Summation of segmental forces over the intruder gives the net forces on the intruder. This approach is valid when the intruder moves slowly enough (< 50 cm/s, below which particle inertial force only contributes to < 10% of the total force) such that granular forces are dominated by friction (speed-independent) and the inertia of the accelerated particles is small [14].

To develop a resistive force model to predict ground reaction forces for legged locomotion on granular media (Fig. 1), we considered an intruder of complex morphology and kinematics moving within the vertical plane, e.g., a bio-inspired L-shaped leg (L-leg) of a hexapedal robot (Fig. 1a) rotated through granular media during locomotion. At each instant of time during rotation, each leg segment (gray bars in Fig. 1b) has a particular depth, orientation, and direction of motion (arrows in Fig. 1b), which may change at the next instant.

Therefore, we measured segmental lift $f_z$ and drag $f_x$ as a function of depth $|z|$, angle of attack $\beta$, and angle of intrusion $\gamma$ (Fig. 1c). We used a robotic arm to move a limb "segment", a small aluminum plate of area $A = 3.8 \times 2.5$ cm$^2$, 0.6 cm thick, at a slow speed of 1 cm/s, as a force sensor measured forces during both intrusion and extraction. The granular medium (~ 1 mm diameter poppy seeds) was prepared by a fluidized bed into a uniform, repeatable, loosely packed state (volume fraction = 0.58) before each intrusion and extraction [7].

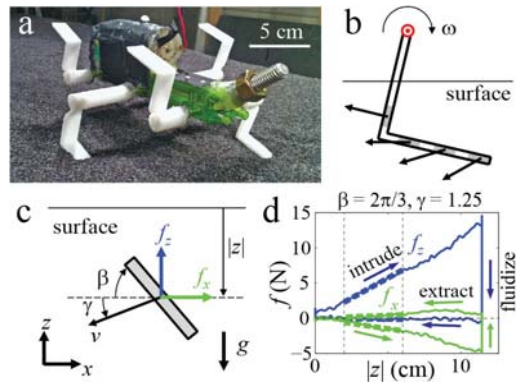

Figure 1. Development of the resistive force model for legged locomotion on granular media. (a) A hexapedal robot with L-legs standing on a granular substrate (~1 mm poppy seeds). (b) Representative leg segments (gray) have various depth, orientation, and direction of motion (arrows) as the L-leg rotates about the hip (red) through granular media within the vertical plane. (c) Segmental lift $f_z$ and drag $f_x$ on a plate depend on depth $|z|$, angle of attack $\beta$, and angle of intrusion $\gamma$. (d) $f_z$ and $f_x$ vs. $|z|$ for representative $\beta = 2\pi/3$ and $\gamma = 1.25$.

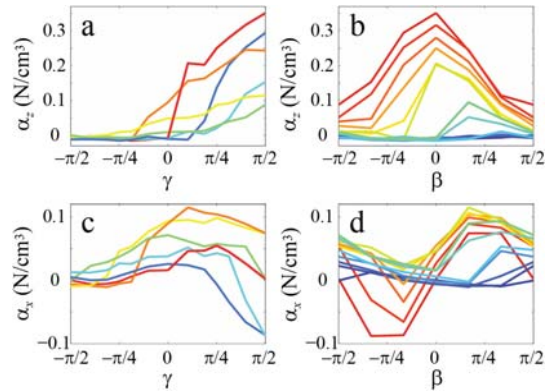

Figure 2. Vertical stress $\alpha_z$ (a,b) and horizontal stresses $\alpha_x$ (c,d) as a function of angle of intrusion $\gamma$ and angle of attack $\beta$. In (a,c), colors from red to blue correspond to $\beta = 0$, 0.52, 1.05, $\pi/2$, 2.09, 2.62, and $\pi$. In (b,d), colors from red to blue correspond to $\gamma = \pi/2$, 1.25, 0.98, $\pi/4$, 0.59, 0.32, 0, $-0.32$, $-0.59$, $-\pi/4$, $-0.98$, $-1.25$, and $-\pi/2$.

We found that for all β and γ, both $f_z$ and $f_x$ were approximately proportional to |z| during both intrusion and extraction (Fig. 1d). We performed a linear fit with zero intercept, $f_{z,x} = \alpha_{z,x} A |z|$ (where $A$ is the segment area), to segmental forces vs. depth data over the steady state regions (dashed lines in Fig. 1d) to determine vertical and horizontal stresses $\alpha_{z,x}$ as a function of β and γ. Both $\alpha_z$ and $\alpha_x$ depended sensitively on β and γ (Fig. 2). For all β, both $\alpha_z$ and $\alpha_x$ had larger magnitudes (absolute value) for positive γ than for negative γ (Fig. 2a,c); i.e. it was generally harder to push the plate into granular media than to extract the plate. For all γ but γ = 0 and π/2, the magnitudes of both $\alpha_z$ and $\alpha_x$ were asymmetric about β = 0 (Fig. 2b,d); i.e., to move a plate with vertically mirroring orientations along the same direction generally required different forces. These asymmetries in stresses are due to the symmetry breaking by gravity in the vertical plane and the finite yield stress of granular media [15].

## 3. Predictive power of resistive force model

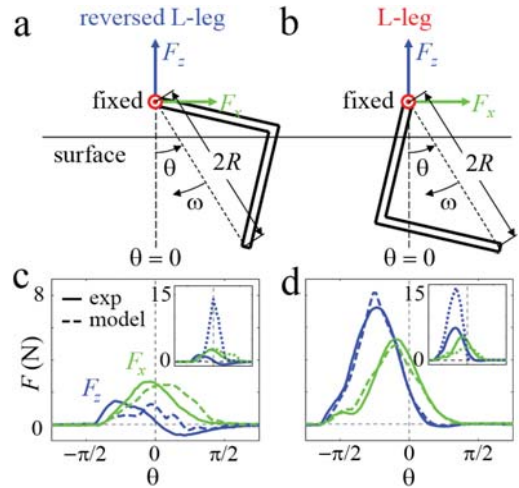

Figure 3. Predictive power of the resistive force model. A reversed L-leg (a) and an L-leg (b) were rotated about a fixed axle (red) through granular media as net lift $F_z$ and drag $F_x$ were measured as a function of leg angle θ (c,d). Solid and dashed curves in (c,d) are experimental measurements and resistive force model predictions. Insets: comparison of predictions from one-dimensional vertical penetration and horizontal drag force models (dotted) with experimental measurements (solid).

To test the power of the resistive force model for predicting net forces on an intruder of complex morphology and kinematics, we examined the net lift $F_z$ and drag $F_x$ on the an L-leg (2R = 7.6 cm, width = 2.5 cm, made of aluminum) rotated about a fixed axle through granular medium, in both normal and reversed configurations (Fig. 3a,b). We used the robotic arm to perform rotation at angular velocity = 0.2 rad/s (tip speeds ~ 1 cm/s) as the force sensor measured forces as a function of leg angle $\theta$ [11]. The fluidized bed prepared the granular medium to a uniform loosely packed state before each rotation [7]. We calculated $F_z$ and $F_x$ from the resistive force model by summing segmental forces over the intruder using $F_{z,x} = \int \alpha_{z,x}(\beta_s, \gamma_s)|z|_s dA_s$ (where $s$ denotes a segment), and compare them with experimental measurements.

Without any fitting parameters, the model prediction of $F_{z,x}$ vs. $\theta$ on both the L-leg and reversed L-leg (Fig. 3c,d, dashed curves) matched experimental measurements (Fig. 3c,d, solid curves) to within 15%, capturing both the magnitudes and asymmetric profiles. As a comparison, the $F_{z,x}$ vs. $\theta$ predicted by similar integration using the one-dimensional vertical penetration model, i.e., $\alpha_z(\beta, \gamma > 0) = \alpha_z(0, \pi/2)$ and $\alpha_z(\beta, \gamma < 0) = \alpha_z(0, -\pi/2)$ [9], and horizontal drag force model, i.e., $\alpha_z(\beta, \gamma) = \alpha_z(\pi/2, 0)$ [10] (Fig. 3c,d, inset, dotted curves) did not agree: it over-predicted $F_z$ by 800% and under-predicted $F_x$ by 50% for the reversed L-leg, and over-predicted $F_z$ by 200% for the L-leg. The discrepancy in the exact shape of $F_{z,x}$ vs. $\theta$ between experiment and resistive force model for the reversed L-leg may be because particles became contained more easily within a concave leading surface [15].

## 4. Predicting legged locomotor performance

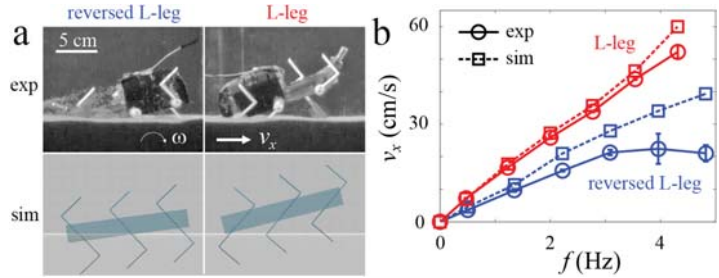

Figure 4. Legged locomotor performance in experiment and simulation. (a) Side views of a hexapedal robot (top) and a simulated robot (bottom) moving on granular media at stride frequency $f \approx 2$ Hz using both reversed L-legs (left) and L-legs (right). (b) Average forward speed $v_x$ as a function of stride frequency $f$. Solid and dashed curves in (b) are experimental measurements and simulation predictions using the resistive force model.

To test whether the resistive force model could be used to predict legged locomotor performance, we studied in both experiments and simulation the average forward speed $v_x$ of a small, bio-inspired, hexapedal robot (15 cm, 150 g) moving on granular media using both L-legs and reversed L-legs (Fig. 4). We used high speed videos to record both dorsal and lateral views as the physical robot ran across granular media (Fig. 4, top). The robot legs were made from plastic with similar friction coefficient (0.36) to that of aluminum (0.40) relative to poppy seeds. The granular medium was prepared into a uniform loosely packed state by a fluidized bed trackway before each run [7]. A multi-body dynamic simulation of the robot was created using MBDyn [16] (Fig. 4a, bottom), which used the resistive force model to calculate ground reaction forces and torques from the robot morphology and kinematics. This simulation required ~ 10 s to simulate 1 s of locomotion. The stride frequency $f$ of the robot was varied between 1 Hz and 5 Hz in both experiment and simulation, and $v_x$ was measured.

The robot in experiment and simulated robot displayed similar kinematics (Fig. 4a) during locomotion. We found that the robot always moved faster at any $f$ with L-legs than with reversed L-legs (Fig. 4b). The simulation prediction of $v_x$ vs. $f$ (Fig. 4b, dashed curves) matched experimental measurements (Fig. 4b, solid curves) to within 15% for all $f$ tested using both L-legs, and to within 30% for reversed L-legs up to $f$ = 3 Hz (Fig. 4b, solid curve). The plateau and larger variation in $v_x$ for $f$ > 3 Hz using the reversed L-leg was a result of the pitch-instability of the robot observed in experiment, which was not found in simulation. As a comparison, similar simulation using the one-dimensional vertical penetration [9] and horizontal drag [10] force models over-predicted $v_x$ vs. $f$ by 50% for the L-leg and by 100% for the reversed L-leg.

## 5. Leg shape variation test

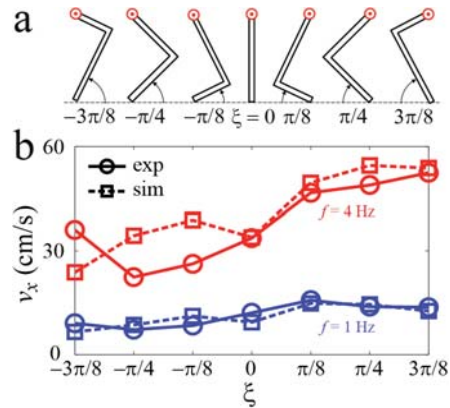

Figure 5. Leg shape variation test. (a) Definition of the angle $\xi$ which parameterizes the shape of the L-leg. (b) Average forward speed $v_x$ vs. $\xi$ for two representative stride frequencies, $f$ = 1 and 4 Hz.

Solid and dashed curves in (b) are experimental measurements (interpolated from data in Fig. 4b) and simulation predictions using the resistive force model.

With the resistive force model and multi-body dynamic simulation, we can quickly perform design and performance tests for legged locomotors on granular media. To demonstrate this capability, in both experiment and simulation we changed the shape of the L-legs by varying the angle $\xi$ which defines the location of the point adjoining the two straight pieces of an L-leg (Fig. 5a), and examined whether the simulation could predict how changes in leg shape affected the average forward speed $v_x$ of the robot.

We found that $v_x$ was higher using legs of positive $\xi$ than using legs of negative $\xi$ (Fig. 5b). The simulation (Fig. 5b, dashed curves) captured the measured $v_x$ vs. $\xi$ to within 10% for positive $\xi$ and to within 50% for negative $\xi$. In addition, agreement was better (within 30%) at $f = 1$ Hz and worse (within 50%) at $f = 4$ Hz. As a comparison, similar simulation using the one-dimensional vertical penetration [9] and horizontal drag [10] force models over-predicted $v_x$ by up to 200% at $f = 1$ Hz and by up to 100% at $f = 4$ Hz. The larger discrepancy between experiment and simulation using the resistive force model for negative $\xi$ may be a consequence of particles being contained within concave leading surfaces. The larger discrepancy at higher $f$ may be because as the robot legs moved faster, inertial effects became significant. Both these effects are not yet captured by the resistive force model.

## 6. Conclusions

We developed a resistive force model in the vertical plane for legged locomotion on granular media. Without any fitting parameters, the resistive force model demonstrated superior predictive power over previously used one-dimensional vertical penetration and horizontal drag force models in predicting the net forces on an intruder with complex morphology and kinematics. Furthermore, a dynamic simulation using the resistive force model predicted the locomotor performance of a small legged robot on granular media using L-legs of a variety of shapes, with significantly better accuracy than that using vertical penetration and horizontal drag force models. By contrast, in a previous study where the vertical penetration force model was used to explain the quasi-static locomotion of a similar robot, two fitting parameters were required to obtain agreement between experiment and model.

Our resistive force model provides a promising candidate for a general force model to calculate forces accurately and quickly for legged locomotion on uniform granular media. By contrast, a discrete element method simulation using same granular bed would contain 5 million poppy seed particles and take 30 days to simulate 1 s of locomotion. Aided by experimental and computational tools, the resistive force model promises to accelerate the design and performance tests for future legged robots on granular media.

**Acknowledgments**

We thank Yang Ding, Nick Gravish, Paul Umbanhowar, Gareth Meirion-Griffith, and Hal Komsuoglu for discussion, Jeff Shen for help with robot modification, and Pierangelo Masarati for MBDyn support.

**References**

1. C. Li, S. T. Hsieh, and D. I. Goldman, *J. Exp. Biol.* (in review).
2. J. Kumagai, *IEEE Spectrum* **6**, 44 (2004).
3. R. Pfeifer, M. Lungarella, and F. Iida, *Science* **318**, 1088 (2007).
4. S. Vogel, *Life in Moving Fluids: The Physical Biology of Flow* (1996).
5. M. G. Bekker, *Off-the-road locomotion, research and development in terramechanics* (1960).
6. J. Y. Wong, *Terramechanics and Off-Road Vehicle Engineering* (2010).
7. C. Li, P. B. Umbanhowar, H. Komsuoglu, D. E. Koditschek, and D. I. Goldman, *Proc. Natl. Acad. Sci.* **106**, 3029 (2009).
8. N. Mazouchova, N. Gravish, A. Savu, and D. I. Goldman, *Biol. Lett.* **6**, 398 (2010).
9. G. Hill, S. Yeung, and S. A. Koehler, *Europhys. Lett.* **72**, 137 (2005).
10. N. Gravish, P. B. Umbanhowar, and D. I. Goldman, *Phys. Rev. Lett.* **105**, 208301 (2010).
11. C. Li, P. B. Umbanhowar, H. Komsuoglu, and D. I. Goldman, *Exp. Mech.* **50**, 1383 (2010).
12. C. Li, A. M. Hoover, P. Birkmeyer, P. B. Umbanhowar, R. S. Fearing, and D. I. Goldman, *Proc. SPIE* **7679**, 1 (2010).
13. R. Maladen, Y. Ding, P. B. Umbanhowar, A. Kamor, and D. I. Goldman, *J. Roy. Soc. Int.* **8**, 1332, (2011).
14. R. D. Maladen, Y. Ding, C. Li, and D. I. Goldman, *Science* **325**, 314 (2009).
15. Y. Ding, N. Gravish, and D. I. Goldman, *Phys. Rev. Lett.* **106**, 028001 (2010).
16. G. Ghiringhelli, P. Masarati, and P. Mantegazza, *Nonlinear Dynamics* **19**, 333 (1999).